\documentclass[11pt]{article} 
\usepackage{times}
\input epsf.sty

\def \lket {|}
\def \rket {\rangle}
\def \lbra {\langle}

\newcommand{\ket}[1]{\lket #1\rket}

\newtheorem{Corollary}{Corollary}
\newtheorem{Definition}{Definition}
\newtheorem{Theorem}{Theorem}
\newtheorem{Lemma}{Lemma}
\newtheorem{Claim}{Claim}
\newcommand{\proof}{\noindent {\bf Proof: }}
\newcommand{\qed}{}
\newcommand{\comment}[1]{}

\begin{document}

\title{Polynomial degree vs. quantum query complexity}
\date{}

\author{
Andris Ambainis\thanks{
Department of Combinatorics and Optimization and Institute for Quantum 
Computing, University of Waterloo, 200 University Avenue West,
Waterloo, ON N2L 2T2, Canada,
e-mail:{\tt ambainis@math.uwaterloo.ca}.
Supported by IQC University Professorship and CIAR.
This work done at Institute of Mathematics and Computer Science, University of Latvia,
Raina bulv. 19,
R\=\i ga, LV-1459, Latvia, supported in part by Latvia Science Council Grant 01.0354}}

\maketitle
\thispagestyle{empty}

\begin{abstract}
The degree of a polynomial representing (or approximating)
a function $f$ is a lower bound for the quantum query complexity
of $f$. This observation has been a source of 
many lower bounds on quantum algorithms. It has been
an open problem whether this lower bound is tight.

We exhibit a function with polynomial degree $M$ and quantum
query complexity $\Omega(M^{1.321...})$. This is
the first superlinear separation between polynomial degree
and quantum query complexity. The lower bound is shown
by a generalized version of the quantum adversary method.
\end{abstract}

\section{Introduction}

Quantum computing provides speedups for 
factoring \cite{Shor}, search \cite{Grover} and many related
problems. These speedups can be quite surprising.
For example, Grover's search algorithm \cite{Grover} 
solves an arbitrary exhaustive search problem with $N$ possibilities
in time $O(\sqrt{N})$. Classically, it is 
obvious that time $\Omega(N)$ would be needed.

This makes lower bounds particularly important in
the quantum world. If we can search in time $O(\sqrt{N})$,
why can we not search in time $O(\log^c N)$? 
(Among other things, that would have meant $NP\subseteq BQP$.)
Lower bound of Bennett et al. \cite{BBBV} shows that this is not possible
and Grover's algorithm is exactly optimal.

\comment{In the classical world, we can 
rely on our intuition on what might be possible and what should
be impossible. In the quantum world for which we do not have 
good intuition, only lower bounds can show that something
is impossible.}

Currently, we have good lower bounds on the quantum complexity
of many problems.
They mainly follow by two methods\footnote{Other approaches,
such as reducing query complexity to communication complexity \cite{BCW}
are known, but have been less successful.}:
the hybrid/adversary method\cite{BBBV,Ambainis0} and
the polynomials method \cite{BB+}.
The polynomials method is useful for proving lower bounds
both in classical \cite{NS} and quantum complexity \cite{BB+}. 
It is known that 
\begin{enumerate}
\item
the number of queries $Q_E(f)$ needed 
to compute a Boolean function $f$ by an exact quantum algorithm
exactly is at least $\frac{\deg(f)}{2}$, where $\deg(f)$ is the degree of
the multilinear polynomial representing $f$,       
\item
the number of queries $Q_2(f)$ needed to 
compute $f$ by a quantum algorithm
with two-sided error is at least $\frac{\widetilde{\deg(f)}}{2}$, 
where $\widetilde{\deg(f)}$ is the smallest degree of a
multilinear polynomial approximating $f$.      
\end{enumerate}
This reduces proving lower bounds on quantum algorithms
to proving lower bounds on degree of polynomials.
This is a well-studied mathematical problem with
methods from approximation theory \cite{ApprThy} available.
Quantum lower bounds shown by polynomials method 
include a $Q_2(f)=\Omega(\sqrt[6]{D(f)})$ relation for
any total Boolean function $f$ \cite{BB+}, lower bounds
on finding mean and median \cite{NW}, collisions and element
distinctness \cite{AS,Kutin}.
Polynomials method is also a key part of recent $\Omega(\sqrt{N})$
lower bound on set disjointness which resolved a longstanding open
problem in quantum communication complexity \cite{Razborov}.

Given the usefulness of polynomials method, it is
an important question how tight is the polynomials
lower bound. \cite{BB+,BWSurvey} proved that, for
all total Boolean functions, $Q_2(f)=O(\deg^6(f))$
and $Q_E(f)=O(\deg^4(f))$. 
The second result was recently improved to $Q_E(f)=O(\deg^3(f))$ 
\cite{Midrijanis}. 
Thus, the bound is tight
up to polynomial factor.

Even stronger result would be 
$Q_E(f)=O(\deg(f))$ or $Q_2(f)=O(\widetilde{\deg(f)})$.
Then, determining the quantum complexity would be
equivalent to determining the degree
of a function as a polynomial.
It has been an open problem to prove or disprove
either of these two equalities \cite{BB+,BWSurvey}.

In this paper, we show the first provable gap 
between polynomial degree and quantum complexity:
$\deg(f)=2^d$ and $Q_2(f)=\Omega(2.5^d)$.
Since $\deg(f)\geq \widetilde{\deg(f)}$ and $Q_E(f)\geq Q_2(f)$,
this implies a separation both between $Q_E(f)$ and
$\deg(f)$ and between $Q_2(f)$ and $\widetilde{\deg}(f)$.

To prove the lower bound, we use the quantum adversary 
method of \cite{Ambainis0}. The quantum adversary method runs a quantum
algorithm on different inputs from some set.
If every input in this set can be changed in many
different ways so that the value of the function
changes, many queries are needed.

The previously known version of quantum adversary method
gives a weaker lower bound of $Q_2(f)=\Omega(2.1213...^d)$.
While this already gives some gap between polynomial degree 
and quantum complexity, we can achieve a larger gap by using a
new, more general version of the method.

The new component is that we carry out this
argument in a very general way.
We assign individual weights to
every pair of inputs and distribute each weight
among the two inputs in an arbitrary way.
This allows us to obtain better bounds than 
with the previous versions of the quantum adversary method.

We apply the new lower bound theorem to three functions
for which deterministic query complexity
is significantly higher than polynomial degree.
The result is that, for all of those functions,
quantum query complexity is higher than polynomial
degree. The biggest gap is polynomial degree $2^d=M$
and query complexity $\Omega(2.5^d)=\Omega(M^{1.321...})$.

Spalek and Szegedy \cite{Szegedy1} have recently shown
that our method is equivalent to two other methods,
the spectral method of \cite{BSS} that was known prior
to our work and the Kolmogorov complexity method of 
\cite{Laplante} that appeared after the conference
version of our paper was published. 
Although all three methods are equivalent,
they have different intuition.
It appears to us that our method is the easiest to use
for results in this paper.

\section{Preliminaries}

\subsection{Quantum query algorithms}

Let $[N]$ denote $\{1, \ldots, N\}$.

We consider computing a Boolean function
$f(x_1, \ldots, x_N):\{0, 1\}^N\rightarrow \{0, 1\}$
in the quantum query model (for a survey on query model, 
see \cite{ASurvey,BWSurvey}).
In this model, the input bits can be accessed by queries to an oracle $X$
and the complexity of $f$ is the number of queries needed to compute $f$.
A quantum computation with $T$ queries
is just a sequence of unitary transformations
\[ U_0\rightarrow O\rightarrow U_1\rightarrow O\rightarrow\ldots
\rightarrow U_{T-1}\rightarrow O\rightarrow U_T.\]

The $U_j$'s can be arbitrary unitary transformations that do not depend
on the input bits $x_1, \ldots, x_N$. The $O$'s are query (oracle) transformations
which depend on $x_1, \ldots, x_N$.
To define $O$, we represent basis states as $|i, z\rangle$ where
$i$ consists of $\lceil \log (N+1)\rceil$ bits and
$z$ consists of all other bits. Then, $O_x$ maps
$\ket{0, z}$ to itself and
$\ket{i, z}$ to $(-1)^{x_i}\ket{i, z}$ for $i\in\{1, ..., N\}$ 
(i.e., we change phase depending on $x_i$, unless $i=0$ in which case we do
nothing). 

The computation starts with a state $|0\rangle$.
Then, we apply $U_0$, $O_x$, $\ldots$, $O_x$,
$U_T$ and measure the final state.
The result of the computation is the rightmost bit of
the state obtained by the measurement.

The quantum computation computes 
$f$ exactly if, for every $x=(x_1, \ldots, x_N)$,
the rightmost bit 
of $U_T O_x \ldots O_x U_0\ket{0}$ 
equals $f(x_1, \ldots, x_N)$ with certainty.

The quantum computation computes
$f$ with bounded error if, for every $x=(x_1, \ldots, x_N)$,
the probability that the rightmost bit 
of $U_T O_x U_{T-1} \ldots O_x U_0\ket{0}$ 
equals  $f(x_1, \ldots, x_N)$ is at
least $1-\epsilon$ for some fixed $\epsilon<1/2$.

$Q_E(f)$ ($Q_2(f)$) denotes the minimum number $T$ of queries
in a quantum algorithm that computes $f$ exactly (with
bounded error). $D(f)$ denotes the minimum number of queries
in a deterministic query algorithm computing $f$.

\subsection{Polynomial degree and related quantities}

For any Boolean function $f$, there is a unique multilinear
polynomial $g$ such that $f(x_1, \ldots, x_N)=g(x_1, \ldots, x_N)$
for all $x_1, \ldots, x_N\in\{0, 1\}$.
We say that $g$ {\em represents} $f$.
Let $\deg(f)$ denote the degree of polynomial representing $f$.

A polynomial $g(x_1, \ldots, x_N)$ approximates
$f$ if $1-\epsilon \leq g(x_1, \ldots, x_N) \leq 1$ whenever 
$f(x_1, \ldots, x_N)=1$ and
$0 \leq g(x_1, \ldots, x_N) \leq \epsilon$ whenever 
$f(x_1, \ldots, x_N)=0$.
Let $\widetilde{\deg(f)}$ denote the minimum degree of a polynomial approximating $f$.
It is known that

\begin{Theorem}
\cite{BB+}
\begin{enumerate}
\item
$Q_E(f)= \Omega(\deg(f))$;
\item
$Q_2(f)= \Omega(\widetilde{\deg(f)})$;
\end{enumerate}
\end{Theorem}

This theorem has been a source of many lower bounds on quantum
algorithms \cite{BB+,NW,AS}.

Two other relevant quantities are {\em sensitivity} and {\em block sensitivity}.
The sensitivity of $f$ on input $x=(x_1, \ldots, x_N)$ is just the
number of $i\in[N]$ such that changing the value of 
$x_i$ changes 
the value of $f$:
\[ f(x_1, \ldots, x_N)\neq f(x_1, \ldots, x_{i-1}, 1-x_i, x_{i+1}, \ldots, x_N) .\]
We denote it $s_x(f)$.
The sensitivity of $f$ is the maximum of $s_x(f)$
over all $x\in\{0, 1\}^N$.
We denote it $s(f)$.

The block sensitivity is a similar quantity in which we flip sets
of variables instead of single variables. For 
$x=(x_1, \ldots, x_N)$ and
$S\subseteq[N]$, let $x^{(S)}$ be the 
input $y$ in which $y_i=x_i$ if $i\notin S$ and $y_i=1-x_i$ if $i\in S$.
The block sensitivity of $f$ on an input $x$ (denoted $bs_x(f)$) is
the maximum number $k$ of pairwise disjoint 
$S_1$, $\ldots$, $S_k$ such that
$f(x^{(S_i)})\neq f(x)$. 
The block sensitivity of $f$ is the maximum of $bs_x(f)$ over all $x\in\{0, 1\}^N$.
We denote it $bs(f)$.

\section{Main results}

\subsection{Overview}
\label{sec:4d}

{\bf The basis function.}
$f(x)$ is equal to 1 iff $x=x_1x_2x_3x_4$ is one
of the following values:
0011, 0100, 0101, 0111, 1000, 1010, 1011, 1100.
This function has the degree of 2, as witnessed by polynomial
$f(x_1, x_2, x_3, x_4)=x_1+x_2+x_3x_4-x_1x_4-x_2x_3-x_1x_2$
and the deterministic complexity $D(f)=3$, as shown in section \ref{sec:lem3}
where we discuss the function in more detail.

\noindent
{\bf Iterated function.}
Define a sequence 
$f^1=f$, $f^2$, $\ldots$ with $f^d$ being a function
of $4^d$ variables by
\[
f^{d+1}=f(f^d(x_1, \ldots, x_{4^d}), 
f^d(x_{4^d+1}, \ldots, x_{2\cdot4^d}), 
\]
\begin{equation}
\label{eq1} 
f^d(x_{2\cdot4^d+1}, \ldots, x_{3\cdot4^d}), 
f^d(x_{3\cdot4^d+1}, \ldots, x_{4^{d+1}})) .
\end{equation}
Then, $deg(f^d)=2^d$, $D(f^d)=3^d$ and,
on every input $x$,
$s_x(f^d)=2^d$ and $bs_x(f^d)=3^d$.

We will show
\begin{Theorem}
\label{GapThm}
$Q_2(f^d)=\Omega(2.5^d)$.
\end{Theorem}

Thus, the exact degree is $deg(f^d)=2^d$ but even the quantum
complexity with 2-sided error $Q_2(f^d)$ is $\Omega(2.5^d)=
deg(f^d)^{1.321..}$. This implies an $M$-vs.-$\Omega(M^{1.321...})$
gap both between exact degree and exact quantum complexity
and between approximate degree and bounded-error quantum complexity.

The proof is by introducing a combinatorial
quantity $Q'_2(f)$ with the following 
properties:

\begin{Lemma}
\label{Lem1}
For any Boolean function $g$, $Q_2(g)=\Omega(Q'_2(g))$.
\end{Lemma}

\begin{Lemma}
\label{Lem2}
Let $g$ be an arbitrary Boolean function. 
If $g^1$, $g^2$, $\ldots$ is obtained by
iterating $g$ as in equation (\ref{eq1}),
then
\[ Q'_2(g^d)\geq (Q'_2(g))^d .\]
\end{Lemma}

\begin{Lemma}
\label{Lem3}
$Q'_2(f)\geq 2.5$.
\end{Lemma}

Theorem \ref{GapThm} then follows from Lemmas \ref{Lem1},
\ref{Lem2}, \ref{Lem3}.

\subsection{Previous methods}

\noindent
Our approach is a generalization of the quantum adversary 
method \cite{Ambainis0}. 

\begin{Theorem}
\label{AThm2}
\cite{Ambainis0}
Let $A\subset \{0, 1\}^N$, $B\subset\{0, 1\}^N$, 
$R\subset A\times B$
be such that $f(A)=0$, $f(B)=1$ and 
\begin{itemize}
\item
for every $x\in A$, there are at least $m$ inputs
$y\in B$ such that $(x, y)\in R$,
\item
for every $y\in B$, there are at least $m'$ inputs
$x\in A$ such that $(x, y)\in R$,
\item
for every $x=(x_1\ldots x_N)\in A$ and every $i\in[N]$
there are at most $l$ inputs $y\in B$ such that $(x, y)\in R$
and $x_i\neq y_i$,
\item
for every $y=(y_1\ldots y_N)\in B$ and every $i\in[N]$,
there are at most $l'$ inputs $x\in A$ such that $(x, y)\in R$
and $x_i\neq y_i$.
\end{itemize}
Then, $Q_2(f)=\Omega(\sqrt{\frac{m m'}{l l'}})$.
\end{Theorem}

\comment{
It might look that this theorem gives $Q'_2(f)=\Omega(3^d)$
because the block sensitivity is $3^d$ on every input.
However, this is not the case. }

There are several ways to apply this theorem to $f^d$
defined in the previous section. 
The best lower bound that can be obtained by it 
seems to be 
$Q_2(f)=\Omega(2.1213..^d)$ (cf. appendix \ref{app:old}).
\comment{
This shows some gap between $Q_2(f)$ and $\deg(

\comment{Say, we try to prove a lower bound by setting
$A=f^{-1}(0)$, $B=f^{-1}(1)$.
Then, for every $x\in A$, we would construct $3^d$
non-intersecting blocks $S_1$, $S_{3^d}$
of variables to which it is sensitive
and add $(x, x^{S_j})$ to $R$.
This does achieve $m=3^d$ and $l=1$.
But, to get a lower bound of $3^d$, we also
need $m'=3^d$ and $l'=1$.
And we fail to make $l'=1$.
If $y\in B$ is obtained from $x\in A$ by flipping
all variables in $S$ and from $z\in A$
by flipping all variables in $T$, there is
no reason for $S$ and $T$ to be disjoint.
}

\comment{A more careful look reveals that this
construction gives $m'=3^d$ and $l'=2^d$.
Thus, we get a lower bound of $\Omega(\sqrt{4.5^d})=
\Omega(2.1213..^d)$ for $Q_2(f)$. }}
This gives some separation between $Q_2(f)$ and $deg(f)=2^d$
but is weaker than our new method
that we introduce next.

\subsection{New method: weight schemes}

We now formally define the combinatorial quantity $Q'_2(f)$ that we use in
Lemmas \ref{Lem1}, \ref{Lem2} and \ref{Lem3}.

\begin{Definition}
Let $f:\{0, 1\}^N \rightarrow \{0, 1\}$,
$A\subseteq f^{-1}(0)$, $B\subseteq f^{-1}(1)$ and
$R\subseteq A\times B$.
A weight scheme for $A, B, R$ consists of numbers
$w(x, y)>0$, $w'(x, y, i)>0$, $w'(y, x, i)>0$ 
for all $(x, y)\in R$ and $i\in[N]$
satisfying $x_i\neq y_i$, we have
\begin{equation} 
\label{eqWReq}
w'(x, y, i)w'(y, x, i)\geq w^2(x, y).
\end{equation}
\end{Definition}

\begin{Definition}
The weight of $x$ is  
$wt(x)=\sum_{y:(x, y)\in R}w(x, y)$, if $x\in A$ and
$wt(x)=\sum_{y:(y, x)\in R}w(x, y)$ if $x\in B$.
\end{Definition}

\begin{Definition}
Let $i\in[N]$.
The load of variable $x_i$ in assignment $x$
is 
\[ v(x, i)=\sum_{y: (x, y)\in R, x_i\neq y_i} w'(x, y, i) \]
if $x\in A$ and
\[ v(x, i)=\sum_{y: (y, x)\in R, x_i\neq y_i} w'(x, y, i) \]
if $x\in B$.
\end{Definition}

We are interested in schemes in which the load of each variable is
small compared to the weight of $x$.

Let the maximum A-load be $v_{A}=\max_{x\in A, i\in[N]} \frac{v(x, i)}{wt(x)}$.
Let the maximum B-load be $v_{B}=\max_{x\in B, i\in[N]} \frac{v(x, i)}{wt(x)}$.
The maximum load of a weight scheme is
$v_{max}=\sqrt{v_A v_B}$.

Let $Q'_2(f)$ be the maximum of $\frac{1}{v_{max}}$ 
over all choices of $A\subseteq\{0, 1\}^N$,
$B\subseteq\{0, 1\}^N$, $R\subseteq A\times B$ and 
all weight schemes for $A, B, R$.
We will show in Lemma \ref{Lem1}, if we have a weight
scheme with maximum load $v_{max}$, the query 
complexity has to be $\Omega(\frac{1}{v_{max}})$.

\subsection{Relation to other methods}

Theorem \ref{AThm2} follows from
our new Lemma \ref{Lem1} if we set $w(x, y)=1$
for all $(x, y)\in R$ and 
$w(x, y, i)=w(y, x, i)=1$ for all 
$i\in[N]$.
Then, the weight of $x$ is just the number of pairs $(x, y)\in R$.
Therefore, $wt(x)\geq m$ for all $x\in A$ and
$wt(y)\geq m'$ for all $y\in B$.
The load of $i$ in $x$ is just the number of $(x, y)\in R$
such that $x_i\neq y_i$.
That is, $v(x, i)\leq l$ and $v(y, i)\leq l'$.
Therefore, $v_A\leq \frac{l}{m}$, $v_B\leq \frac{l'}{m'}$ and
$v_{max}\leq \sqrt{\frac{l l'}{m m'}}$.
This gives us the lower bound of Theorem \ref{AThm2}.

There are several generalizations of Theorem \ref{AThm2} 
that have been proposed. Barnum and Saks \cite{BS}
have a generalization of Theorem \ref{AThm2} 
that they use to prove a $\Omega(\sqrt{N})$ lower
bound for any read-once function on $N$ variables.
This generalization can be shown to be a particular
case of our Lemma \ref{Lem1}, with a weight scheme
constructed in a certain way.

Barnum, Saks and Szegedy \cite{BSS} have a very general
and promising approach. They reduce quantum query complexity
to semidefinite programming and show that a $t$-query
algorithm exists if and only if a certain semidefinite 
program does not have a solution. 
Spalek and Szegedy have recently shown \cite{Szegedy1} that our weighted 
scheme method is equivalent to Theorem 4 
in \cite{BSS} which is a special case of their general method.
Our method is also equivalent \cite{Szegedy1} to Kolmogorov 
complexity method by Laplante and Magniez \cite{Laplante}.


Hoyer, Neerbek and Shi \cite{HNS} have shown lower bounds 
for ordered searching and sorting using a weighted version
of the quantum adversary method, before 
both this paper and \cite{BSS}. Their argument can be described
as a weight scheme for those problems, but it is more natural
to think about it in the spectral terminology of \cite{BSS}.

\section{Proofs}

\subsection{Lemma \ref{Lem1}}

In terms of weights schemes, Lemma \ref{Lem1} becomes

\medskip
\noindent
{\bf Lemma \ref{Lem1}}
{\em If a function $g$ has a weight scheme
with maximum load $v_{max}$, 
then $Q_2(g)=\Omega(\frac{1}{v_{max}})$.}
\medskip

\proof
We can assume that $v_A=v_B=v_{max}$.
Otherwise, we just multiply
all $w'(x, y, i)$ by $\sqrt{v_B/v_A}$ and all 
$w'(y, x, i)$ by $\sqrt{v_A/v_B}$. 
Notice that this does not affect
the requirement (\ref{eqWReq}). 
In the new scheme $v_A$ is equal to
the old $v_A\sqrt{v_B/v_A}=\sqrt{v_A v_B}=v_{max}$
and $v_B$ is equal to 
the old $v_B\sqrt{v_A/v_B}=\sqrt{v_A v_B}=v_{max}$.

Let $\ket{\psi_{x}^t}$ be the state of a quantum algorithm
after $t$ queries on input $x$.
We consider 
\[ W_t= \sum_{(x, y)\in R} w(x, y)
|\lbra \psi_{x}^t |\psi_{y}^t\rket| .\]
For $t=0$, $W_0=\sum_{(x, y)\in R} w(x, y)$.
Furthermore, if an algorithm computes 
$f$ in $t$ queries with probability
at least $1-\epsilon$, 
$W_t\leq 2\sqrt{\epsilon(1-\epsilon)} W_0$
\cite{Ambainis0,HNS}.
To prove that $T=\Omega(\frac{1}{v_{max}})$, it
suffices to show

\begin{Lemma}
$|W_j-W_{j-1}|\leq 2 v_{max} W_0$.
\end{Lemma}

\proof
Let $\ket{\phi_{x}^t}$ be the state of the algorithm
immediately before query $t$.
We write 
\[ \ket{\phi_{x}^t} = \sum_{i=0}^N
\alpha^t_{x, i}\ket{i}\ket{\phi'_{x, i}} \]
with $\ket{\phi'_{x, i}}$ being the state
of qubits not involved in the query.
The state after the query is 
\[ \ket{\psi_{x}^t} = \alpha^t_{x, 0}\ket{0}\ket{\phi'_{x, 0}} + \sum_{i=1}^N
\alpha^t_{x, i} (-1)^{x_i} \ket{i}\ket{\phi'_{x, i}} .\]
Notice that all the terms in 
$\lbra \phi_{x}^t |\phi_{y}^t\rket$ and 
$\lbra \psi_{x}^t |\psi_{y}^t\rket$ are the same,
except for those which have $x_i\neq y_i$.
Thus, 
\[ \lbra \psi_{x}^t |\psi_{y}^t\rket - 
\lbra \phi_{x}^t |\phi_{y}^t\rket \leq 
2 \sum_{i:x_i \neq y_i} |\alpha^t_{x, i}||\alpha^t_{y, i}| \]
and 
\[ |W_j-W_{j-1}| \leq  
2 \sum_{(x, y)\in R} \sum_{i:x_i \neq y_i} 
w(x, y) |\alpha^t_{x, i}||\alpha^t_{y, i}|  .\]
By the inequality $2AB\leq A^2+B^2$, 
\[ |W_j-W_{j-1}| \leq  
\sum_{(x, y)\in R} \sum_{i:x_i \neq y_i} 
(w'(x, y, i) |\alpha^t_{x, i}|^2 +
w'(y, x, i) |\alpha^t_{y, i}|^2) .\]
We consider the sum of all first and all second terms separately.
The sum of all first terms is
\[ \sum_{(x, y)\in R} \sum_{i:x_i \neq y_i} 
w'(x, y, i) |\alpha^t_{x, i}|^2  = 
\sum_{x\in A, i\in[N]} |\alpha^t_{x, i}|^2 
\left( \sum_{y:(x, y)\in R, x_i\neq y_i}
w'(x, y, i) \right) \] 
\[= \sum_{x\in A, i\in[N]} |\alpha^t_{x, i}|^2 v(x, i)
\leq v_A \sum_{x\in A, i\in [N]} |\alpha^t_{x, i}|^2 wt(x) \]
\[=v_A \sum_{x\in A} wt(x) \sum_{i\in [N]} |\alpha^t_{x, i}|^2 =
v_A \sum_{x\in A} wt(x) = v_A W_0 .\]
Similarly, the second sum is at most $v_B W_0$.
Finally, $v_A=v_B=v_{max}$ implies that
$|W_j-W_{j-1}|\leq 2v_{max}W_0$.
\qed
\qed

\subsection{Lemma \ref{Lem2}}
\label{sec:lem2}

In terms of weight schemes, we have to prove

\medskip
\noindent
{\bf Lemma \ref{Lem2}}
{\em
Let $g$ be a function with a weight scheme
with maximum load $v_{1}$.
Then, the function $g^d$ obtained by iterating $g$
as in equation (\ref{eq1}) has a weight scheme
with maximum load $v_{1}^d$.}
\medskip

The lemma follows by inductively applying

\begin{Lemma}
If $g$ has a weight scheme with maximum load
$v_1$ and $g^{d-1}$ has a weight scheme with
maximum load $v_{d-1}$, then $g^d$
has a weight scheme with maximum load $v_1v_{d-1}$.
\end{Lemma}

\proof
Similarly to 
lemma \ref{Lem1}, assume that the schemes for $g$ and $g^{d-1}$
have $v_A=v_B=v_{max}$.

Let $n$ be the number of variables for
the base function $g(x_1, \ldots, x_n)$. 
We subdivide the $n^d$ variables $x_1$, $\ldots$, $x_{n^d}$ of 
the function $g^d$ into $n$ blocks of $n^{d-1}$ variables.
Let $x^j=(x_{(j-1)n^{d-1}+1}$, $\ldots$, $x_{jn^{d-1}})$
be the $j^{\rm th}$ block.
Furthermore, let $\tilde{x}$ be the vector
\[ (g^{d-1}(x^1), g^{d-1}(x^2), \ldots
g^{d-1}(x^n)) .\]
Then, $g^d(x)=g(\tilde{x})$.

We start by defining $A$, $B$ and $R$.
Let $A_1$, $B_1$, $R_1$ ($A_{d-1}$,
$B_{d-1}$, $R_{d-1}$) be $A$, $B$, $R$ in
the weight scheme for $g$ ($g^{d-1}$, respectively).
$x\in A$ ($B$, respectively) if 
\begin{itemize}
\item
$\tilde{x}\in A_1$ ($B_1$, respectively), and
\item
for every $j\in[n]$,
$x^j\in A_{d-1}$ if $\tilde{x}_j=0$ and 
$x^j\in B_{d-1}$ if $\tilde{x}_j=1$.
\end{itemize}
$(x, y)\in R$ if $(\tilde{x}, \tilde{y})\in R_1$ and,
for every $j\in[n]$,
\begin{itemize}
\item
$x^j=y^j$ if $\tilde{x}_j=\tilde{y}_j$.
\item
$(x^j, y^j)\in R_{d-1}$ if $\tilde{x}_j=0$,
$\tilde{y}_j=1$.
\item
$(y^j, x^j)\in R_{d-1}$ if $\tilde{x}_j=1$,
$\tilde{y}_j=0$.
\end{itemize}

Let $w_1(x, y)$ denote the weights in the scheme
for $g$ and $w_{d-1}(x, y)$ the weights in the scheme
for $g^{d-1}$. We define the weights 
for $g^d$ as
\[ 
w_d(x, y)=w_1(\tilde{x}, \tilde{y})
\prod_{j: \tilde{x}_j = \tilde{y}_j} wt_{d-1}(x^j)
\prod_{j: \tilde{x}_j \neq \tilde{y}_j} 
w_{d-1}(x^j, y^j) \]
where $wt_{d-1}$ is the weight of 
$x^j$ in the scheme for $g^{d-1}$.

For $i\in[n^d]$, let 
$i_1=\lceil \frac{i}{n^{d-1}}\rceil$ 
be the index of the block containing $i$
and $i_2=(i-1)\bmod n^{d-1} +1$
be the index of $i$ within this block.
Define
\[ w'_d(x, y, i)=
w_d(x, y) \sqrt{\frac{w'_{1}(\tilde{x}, \tilde{y}, i_1)}{w'_{1}(\tilde{y}, \tilde{x}, i_1)}}
\sqrt{\frac{w'_{d-1}(x^{i_1}, y^{i_1}, i_2)}{w'_{d-1}(y^{i_1}, x^{i_1}, i_2)}} .\]

The requirement (\ref{eqWReq}) is obviously satisfied.
It remains to show that the maximum load is at most $v_1v_{d-1}$.
We start by calculating the total weight $wt_d(x)$.
First, split the sum of all $w_d(x, y)$
into sums of $w_d(x, y)$ over $y$ with a fixed $z=\tilde{y}$.

\begin{Claim}
\label{ClaimBreakUpSum}
\[ \sum_{y\in\{0, 1\}^{n^d}:\tilde{y}=z} w_d(x, y) =
 w_1(\tilde{x}, z) \prod_{j=1}^n wt_{d-1}(x^j) . \]
\end{Claim}

\proof
Let $y$ be such that $\tilde{y}=z$.
Then,
\[ w_d(x, y)=w_1(\tilde{x}, z)
\prod_{j: \tilde{x}_j = z_j} wt_{d-1}(x^j)
\prod_{j: \tilde{x}_j \neq z_j} w_{d-1}(x^j, y^j) \]
When $\tilde{x}_j \neq z_j$, $y^j$ can be equal to
any $y'\in\{0, 1\}^{n^{d-1}}$ such that $g^{d-1}(y')=z_j$. 
Therefore, the sum of all $w_d(x, y)$, $\tilde{y}=z$ is
\begin{equation}
\label{EqBreakUpSum} 
w_1(\tilde{x}, z)
\prod_{j: \tilde{x}_j = z_j} wt_{d-1}(x^j)\cdot
\prod_{j: \tilde{x}_j \neq z_j} 
\left(\sum_{y'\in\{0, 1\}^{n^{d-1}}: 
g^{d-1}(y')=z_j} w_{d-1}(x^j, y') \right) .
\end{equation}
Each of sums in brackets is equal to
$wt_{d-1}(x^j)$. Therefore, (\ref{EqBreakUpSum}) 
equals
\[ w_1(\tilde{x}, z) \prod_{j=1}^n wt_{d-1}(x^j) .\]
\qed

\begin{Corollary}
\begin{equation}
\label{EqSumInvX} 
wt_{d}(x)= wt_1(\tilde{x}) \prod_{j=1}^n wt_{d-1}(x^j) .
\end{equation}
\end{Corollary}

\proof
$wt_{d}(x)$ is the sum of sums from Claim \ref{ClaimBreakUpSum}
over all $z\in\{0, 1\}^n$. 
Now, the corollary follows from Claim \ref{ClaimBreakUpSum} and 
$\sum_{z\in\{0, 1\}^n} w_1(\tilde{x}, z)=wt_1(\tilde{x})$
(which is just the definition of $wt_1(\tilde{x})$).
\qed

Next, we calculate the load 
\[ v(x, i)=\sum_{y\in\{0, 1\}^{n^d}} w'_d(x, y, i) \] 
in a similar way.
We start by fixing $z=\tilde{y}$ and
all variables in $y$ outside the $i_1^{\rm th}$ block.
Let $W$ be the sum of $w_d(x, y)$ and $V$ be the sum of 
$w'_d(x, y, i)$, over $y$ that have $\tilde{y}=z$ and 
the given values of variables outside $y^{i_1}$.

\begin{Claim}
\[ V\leq v_{d-1} \sqrt{\frac{w'_1(\tilde{x}, \tilde{y}, i_1)}{
w'_1(\tilde{y}, \tilde{x}, i_1)}} W .\]
\end{Claim}

\proof
Fixing $z$ and the variables outside $y^{i_1}$ fixes all
terms in $w_d(x, y)$, except $w_{d-1}(x^{i_1}, y^{i_1})$.
Therefore, $w_d(x, y)= C w_{d-1}(x^{i_1}, y^{i_1})$
where $C$ is fixed.
This means $W=C wt_{d-1}(x^{i_1})$.
Also, 
\[ w'_d(x, y, i) =C w_{d-1}(x^{i_1}, y^{i_1}) \cdot 
 \sqrt{\frac{w'_{d-1}(x^{i_1}, y^{i_1}, i_2)}{
w'_{d-1}(y^{i_1}, x^{i_1}, i_2)}} \sqrt{\frac{w'_{1}(\tilde{x}, \tilde{y}, i_1)}{w'_{1}(\tilde{y}, \tilde{x}, i_1)}}
.\]
Property (\ref{eqWReq}) of the 
scheme for ($A_{d-1}$, $B_{d-1}$, $R_{d-1}$) implies
\[ w_{d-1}(x^{i_1}, y^{i_1}) 
\sqrt{\frac{w'_{d-1}(x^{i_1}, y^{i_1}, i_2)}{
w'_{d-1}(y^{i_1}, x^{i_1}, i_2)}} \leq w'_{d-1}(x^{i_1}, y^{i_1}, i_2) ,\]
\[ w'_d(x, y, i) \leq C w'_{d-1}(x^{i_1}, y^{i_1}, i_2) 
\sqrt{\frac{w'_{1}(\tilde{x}, \tilde{y}, i_1)}{w'_{1}(\tilde{y}, \tilde{x}, i_1)}} .\]
If we sum over all possible $y^{i_1}\in\{0, 1\}^{n^{d-1}}$, we get
\[ V\leq C v_{d-1}(x^{i_1}, i_2) \sqrt{\frac{w'_{1}(\tilde{x}, \tilde{y}, i_1)}{w'_{1}(\tilde{y}, \tilde{x}, i_1)}} \]
Since $v_{d-1}(x^{i_1}, i_2)\leq v_{d-1} wt_{d-1}(x^{i_1})$, we have
\[ V \leq C v_{d-1} wt_{d-1}(x^{i_1}) \sqrt{\frac{w'_{1}(\tilde{x}, \tilde{y}, i_1)}{w'_{1}(\tilde{y}, \tilde{x}, i_1)}} 
 = v_{d-1} \sqrt{\frac{w'_1(\tilde{x}, \tilde{y}, i_1)}{
w'_1(\tilde{y}, \tilde{x}, i_1)}} W .\]
\qed

We now consider the part of $v(x, i)$ generated by $w'_d(x, y, i)$ with a fixed 
$\tilde{y}$. By the argument above, it is at most 
$v_{d-1} \sqrt{\frac{w'_{1}(\tilde{x}, \tilde{y}, i_1)}{w'_{1}(\tilde{y}, \tilde{x}, i_1)}}$
times the sum of corresponding $w_d(x, y)$.
By Claim \ref{ClaimBreakUpSum}, this sum is 
$w_1(\tilde{x}, z) \prod_{j=1}^n wt_{d-1}(x^j)$.
By summing over all $\tilde{y}$, we get
\[ v(x, i) \leq \sum_{z\in\{0, 1\}^n} v_{d-1} 
\sqrt{\frac{w'_{1}(\tilde{x}, z, i_1)}{w'_{1}(z, \tilde{x}, i_1)}}
w_1(\tilde{x}, z) \prod_{j=1}^n wt_{d-1}(x^j) \]
\begin{equation}
\label{eqNewSum} 
= v_{d-1} \prod_{j=1}^n wt_{d-1}(x^j) 
\sum_{z\in\{0, 1\}^n} \sqrt{\frac{w'_{1}(\tilde{x}, z, i_1)}{w'_{1}(z, \tilde{x}, i_1)}}
w_1(\tilde{x}, z) 
\end{equation}
By property (\ref{eqWReq}), 
$\sqrt{\frac{w'_{1}(\tilde{x}, z, i_1)}{w'_{1}(z, \tilde{x}, i_1)}}
w_1(\tilde{x}, z) \leq w'_1(\tilde{x}, z, i_1)$.
Therefore,
\[ \sum_{z\in\{0, 1\}^n} \sqrt{\frac{w'_{1}(\tilde{x}, z, i_1)}{w'_{1}(z, \tilde{x}, i_1)}}
w_1(\tilde{x}, z) 
 \leq \sum_{z\in\{0, 1\}^n} w'_1(\tilde{x}, z, i_1) = v(\tilde{x}, i_1)
\leq v_1 wt(\tilde{x})\] 
and (\ref{eqNewSum}) is at most 
\[ v_{d-1} \prod_{j=1}^n wt_{d-1}(x^j) v_1 wt(\tilde{x}) = v_1 v_{d-1} wt_d(x) \]
\qed

By induction, 
$v_{d}\leq (v_{1})^d$.
This proves lemma \ref{Lem2}.

\subsection{Lemma \ref{Lem3}}
\label{sec:lem3}

\begin{figure*}
\begin{center}
\epsfxsize=4.5in
\hspace{0in}
\epsfbox{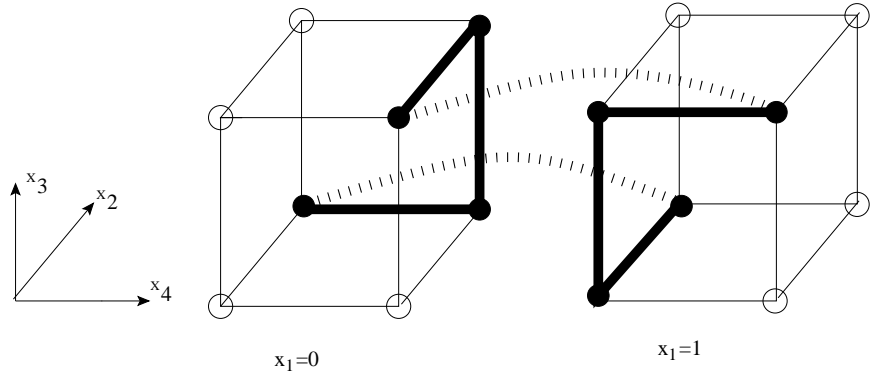}
\caption{The function $f$}
\label{fig-only}
\end{center}
\end{figure*}

We now look at the base function $f$ in more detail.
The function $f$ is shown in Figure \ref{fig-only}.
The vertices of the two cubes correspond to $(x_1, x_2, x_3, x_4)\in\{0, 1\}^4$.
Black circles indicate that $f(x_1, x_2, x_3, x_4)=1$. 
Thick lines connect pairs of black vertices 
that are adjacent (i.e., $x_1x_2x_3x_4$
and $y_1y_2y_3y_4$ differing in exactly one variable with 
$f(x_1, x_2, x_3, x_4)=1$ and $f(y_1, y_2, y_3, y_4)=1$).

From the figure, we can observe several properties.
Each black vertex ($f=1$) has exactly two black neighbors
and two white neighbors.
Each white vertex ($f=0$) also has two white
and two black neighbors.
Thus, for every $x\in\{0, 1\}^4$, 
there are two variables $x_i$ such that changing $x_i$
changes $f(x)$. We call these two 
{\em sensitive} variables and the other two {\em insensitive}.
From figure \ref{fig-only} we also see that, for any $x\in\{0, 1\}^4$,
flipping both sensitive variables changes $f(x)$ and
flipping both insensitive variables also changes $f(x)$.

Thus, the sensitivity of $f$ is 2 on every input. 
The block sensitivity is 3 on every input, with each of the two sensitive 
variables being one block and the two insensitive variables together
forming the third block.

Finally, $D(f)=3$. The algorithm queries
$x_1$ and $x_3$. After both of those are known, the 
function depends only on one of $x_2$ and $x_4$
and only one more query is needed. The lower bound
follows from $bs(f)=3$. 

We now proceed to proving the lemma.
In terms of weight schemes, the lemma is


\medskip
\noindent
{\bf Lemma \ref{Lem3}}
{\em
The function $f$ has a weight scheme with the maximum load of 2.5.
}
\medskip

\proof
Let $A=f^{-1}(0)$, $B=f^{-1}(1)$.
$R$ consists of all $(x, y)$ where $x\in A$ and
$y$ differs from
$x$ in exactly 
\begin{itemize}
\item
one of the sensitive variables or 
\item
both sensitive variables or
\item
both insensitive variables.
\end{itemize}
Thus, for every $x\in A$, there are 
four inputs $y\in B$ such that $(x, y)\in R$.
Also, for every $y\in B$, there are 
four inputs $x\in A$ such that $(x, y)\in R$ and again,
these are $x$ differing from $y$ in 
one sensitive variable, both sensitive variables
or both insensitive variables.
Notice that, if $y$ differs from $x$ in both
variables that are insensitive for $x$,
then those variables are sensitive for $y$
and conversely.
(By flipping one of them in $y$, we get
to an input $z$ which differs from $x$ in
the other variable insensitive to $x$.
Since the variable is insensitive for $x$, $f(x)=f(z)$.
Together with $f(x)\neq f(y)$, this implies $f(y)\neq f(z)$. )

Let $w(x, y)=1$ for $(x, y)\in R$ with $x$, $y$
differing in one variable and $w(x, y)=2/3$
if $x, y$ differ in two variables.
Thus, $wt(x)=2\cdot 1+2\cdot \frac{2}{3}=\frac{10}{3}$ for all $x$.
$w'(x, y, i)$ is 
\begin{itemize}
\item
1 if $x$ and $y$ differ in one variable,
\item
$\frac{1}{3}$ if they differ in both
variables sensitive for $x$,
\item
$\frac{4}{3}$ if they differ in both
variables insensitive for $x$.
\end{itemize}
Since $\frac{1}{3} \cdot\frac{4}{3}=\left(\frac{2}{3}\right)^2$,
this is a correct weight scheme.

We now calculate the load of $i$.
There are two cases.
\begin{enumerate} 
\item
$x$ is insensitive to flipping $x_i$.
Then, the only input $y$ such that $(x, y)\in R$
and $x_i\neq y_i$ is obtained by flipping
both insensitive variables.
It contributes $\frac{4}{3}$ to $v(x, i)$.
\item
$x$ is sensitive to flipping $x_i$.
Then, there are two inputs $y$: one obtained by flipping
just this variable and one obtained by flipping
both sensitive variables. The load is 
$v(x, i)=1+\frac{1}{3}=\frac{4}{3}$.
\end{enumerate} 

Thus, we get $\frac{wt(x)}{v(x, i)}=\frac{10}{4}=
2.5$ for all $x, i$.
\qed

\subsection{Theorem 2}

Theorem 2 now follows from Lemmas \ref{Lem1}, \ref{Lem2}, \ref{Lem3}.
By Lemma \ref{Lem3}, the function $f$ has a weight scheme with
the maximum load of $2.5$. Together with Lemma \ref{Lem2}, this implies
that $f^d$ has a weight scheme with the maximum load of $2.5^d$. 
By Lemma \ref{Lem1}, this means that $Q_2(f)=\Omega(2.5^d)$.

\section{Other base functions}
\label{sec:other1}

Iterated functions similar to ours have been studied before.
Nisan and Wigderson \cite{NW94} used them to show
a gap between communication complexity and log rank 
(an algebraic quantity that provides a lower bound on
communication complexity).
Buhrman and de Wolf \cite{BWSurvey} proposed
to study the functions from \cite{NW94} to find out if polynomial
degree of a function characterizes its quantum complexity.
However, the base functions 
that \cite{NW94,BWSurvey} considered are
different from ours.

We now consider the functions from \cite{NW94,BWSurvey}.
Our method shows the gaps between $\deg(f)$
and $Q_2(f)$ for those functions as well
but those gaps are considerably smaller than for
our new base function.

{\bf Function 1 \cite{NS,NW94}.}
$g(x_1, x_2, x_3)$ is 0 iff all variables are equal.
We have $\deg(g)=2$
(as witnessed by $g=x_1+x_2+x_3-x_1x_2-x_1x_3-x_2x_3$), 
and $D(g)=3$.

\begin{Lemma}
\label{lem:other1}
$g$ has a weight scheme with max load $\sqrt{2}/3$.
\end{Lemma}

\proof
Let $A=g^{-1}(0)$, $B=g^{-1}(1)$, $R=A\times B$.
We set $w(x, y)=2$ if
$x, y$ differ in one variable
and $w(x, y)=1$ if $x$ and $y$ differ into
two variables. (Notice that $x$ and $y$ cannot differ
in all three variables because that would imply $g(x)=g(y)$.)

The total weight $wt(x)$ is 
\begin{itemize}
\item
$3\cdot 2+3\cdot 1=9$ for $x\in A$
(since there are three ways to choose one variable
and three ways to choose two variables and every way of
flipping one or two variables changes the value).
\item
$2+1=3$ for $x\in B$. 
(Each such $x$ has two variables equal and third 
different. It is involved in $w(y, x)$ with
$y$ obtained by flipping either 
the different variable or both equal variables.)
\end{itemize}

Let $x\in A$, $y\in B$.
If $x, y$ differ in one variable $x_i$, we define
$w'(x, y, i)=2\sqrt{2}$ and $w'(y, x, i)=\sqrt{2}$.
If $x, y$ differ in two variables, 
$w'(x, y, i)=\sqrt{2}/2$ and $w'(y, x, i)=\sqrt{2}$
for each of those variables.

The load of $i$ in $x$ is:
\begin{enumerate}
\item
$g(x)=0$.

We have to add up $w'(x, y, i)$ with $y$ differing
from $x$ either in $x_i$ only or in $x_i$ and
one of other two variables. We get 
$2\sqrt{2}+2\cdot(\sqrt{2}/2)=3\sqrt{2}$.
\item
$g(x)=1$.

Then, there is only one input $y$. It can differ in just
$x_i$ or $x_i$ and one more variable. 
In both cases, $w'(x, y, i)=\sqrt{2}$. 
\end{enumerate}
We have
$v_A=\frac{3\sqrt{2}}{9}=\frac{\sqrt{2}}{3}$
and $v_B=\frac{\sqrt{2}}{3}$.
Therefore,
$v_{max}=\frac{\sqrt{2}}{3}$.
\qed

This means that $Q_2(g^d)=\Omega((\frac{3}{\sqrt{2}})^d)=
\Omega(2.12..^d)$.

{\bf Function 2 (Kushilevitz, quoted in \cite{NW94}).}
The function $h(x)$ of 6 variables is defined by
\begin{itemize}
\item $h(x)=0$ if 
the number of $x_i=1$ is 0, 4 or 5, 
\item $h(x)=1$ if 
the number of $x_i=1$ is 1, 2 or 6,
\item if 
the number of $x_i=1$ is 3,
$h(x)=0$ in the following cases: $x_1=x_2=x_3=1$,
$x_2=x_3=x_4=1$, $x_3=x_4=x_5=1$, $x_4=x_5=x_1=1$,
$x_5=x_1=x_2=1$, $x_1=x_3=x_6=1$, $x_1=x_4=x_6=1$,
$x_2=x_4=x_6=1$, $x_2=x_5=x_6=1$, $x_3=x_5=x_6=1$
and 1 otherwise.
\end{itemize}

We have $deg(h)=3$ and $D(h)=6$.

\begin{Lemma}
\label{lem:other2}
$h$ has a weight scheme with max load $4/\sqrt{39}$.
\end{Lemma}

\proof
We choose $A$ to consist of inputs $x$ with all $x_i=0$
and those inputs $x$ with three variables $x_i=1$ which have $h(x)=0$.
$B$ consists of all inputs $x$ with exactly one variable equal to 1.
$R$ consists of $(x, y)$ such that $y$ can be
obtained from $x$ by flipping one variable if $x=0^6$ and
two variables if $x$ contains three $x_i$.

If $x=0^6$ and $y\in B$, we set 
$w(x, y)=w'(x, y, i)=w'(y, x, i)=1$.

If $x$ has three variables $x_i=1$ and
$y$ is obtained by switching two of those to 0,
we set $w(x, y)=1/8$,
$w'(x, y, i)=\frac{1}{32}$ and $w'(y, x, i)=\frac{1}{2}$.

To calculate the maximum loads, we consider three cases:
\begin{enumerate}
\item
$x=0^6$.

$wt(x)=6$ and $v(x, i)=1$ for all $i$.
\item
$x$ has three variables $x_i=1$.

Then, there are three pairs of variables that we can
flip to get to $y\in B$. Thus, $wt(x)=3/8$.
Each $x_i=1$ gets flipped in two of those pairs.
Therefore, its load is $v(x, i)=2\cdot 1/32=1/16$.
The ratio $\frac{wt(x)}{v(x, i)}$ is 6.
\item
$y$ has 1 variable $y_i=1$.

Then, we can either flip this variable or one of 5
pairs of $y_i=0$ variables to get to $x\in A$.
The weight is $wt(y)=1+5\cdot \frac{1}{8}=\frac{13}{8}$.
If $y_i=1$, then the only input $x\in A$, $(x, y)\in R$ 
with $x_i\neq y_i$ is $x=0^6$ with $w'(y, x, i)=1$.
Thus, $v(y, i)=1$.
If $y_i=0$, then exactly
two of 5 pairs of variables $j:y_j=0$ 
include the $i^{\rm th}$ variable.
Therefore, $v(y, i)=2 \cdot \frac{1}{2}=1$.
\end{enumerate}

Thus, $v_A=1/6$, $v_B=8/13$ and
$v_{max}=2/\sqrt{39}$.
\qed

This gives a $3^d$ vs. $\Omega((\sqrt{39}/2)^d)=\Omega(3.12...^d)$
gap between polynomial degree and quantum complexity.

\section{Conclusion}

An immediate open problem is to improve our quantum lower bounds
or to find quantum algorithms for our iterated 
functions that are better
than classical by more than a constant factor. 
Some other related open problems are:
\begin{enumerate}
\item
{\bf AND-OR tree.}
Let
\[ f(x_1, \ldots, x_{4})=
(x_1 \wedge x_2) \vee (x_3 \wedge x_4) .\]
We then iterate $f$ and obtain a function of $N=4^n$ variables 
that can be described by a complete binary tree of depth $\log_2 N=2n$.
The leaves of this tree correspond to variables. 
At each non-leaf node, we take the AND of two values at its two
children nodes at even levels and OR of two values at odd levels.
The value of the function is the value that we get at the root.
Classically, any deterministic algorithm has to query all $N=4^n$ variables.
For probabilistic algorithms, 
$N^{0.753...}=(\frac{1+\sqrt{33}}{4})^{2n}$ queries are 
sufficient and necessary \cite{SW,Santha,Snir}.
What is the quantum complexity of this problem?
No quantum algorithm that uses less than 
$N^{0.753...}=(\frac{1+\sqrt{33}}{4})^{2n}$ queries 
is known but 
the best quantum lower bound is just $\Omega(N^{0.5})=\Omega(2^n)$.

A related problem that has been recently resolved concerns AND-OR trees
of constant depth. There, we have a similar $N^{1/d}$-ary tree of depth $d$.
Then, $\Theta(\sqrt{N})$ quantum queries are sufficient \cite{BCW,HW} and 
necessary \cite{Ambainis0,BS}.
The big-O constant depends on $d$ and the number of queries in the quantum algorithm 
is no longer $O(\sqrt{N})$ if the number of levels is non-constant.
Curiously, it is not known whether the polynomial degree is $\Theta(\sqrt{N})$, even for $d=2$ \cite{Shi1}.
\item
{\bf Certificate complexity barrier.}
Let $C_0(f)$ and $C_1(f)$ be 0-certificate and 1-certificate
complexity of $f$ (cf. \cite{BWSurvey} for definition). 
Any lower bound following from theorems of \cite{Ambainis0} 
or weight schemes of the present paper is $O(\sqrt{C_0(f) C_1(f)})$
for total functions and $O(\sqrt{\min(C_0(f), C_1(f)) N})$ for partial functions\footnote{The distinction between partial 
and total functions is essential here. The methods of \cite{Ambainis0} and the present paper can be used
to prove lower bounds for partial functions that are more than $\sqrt{C_0(f) C_1(f)}$ but $O(\sqrt{\min(C_0(f), C_1(f)) N})$.
Examples are inverting a permutation \cite{Ambainis0} and local search 
\cite{Aaronson1}.}
\cite{Laplante,Zhang}. 

This has been sufficient to prove tight bounds for many functions.
However, in some cases quantum complexity is (or seems to be) higher. 
For example, the binary AND-OR tree described above has $C_0(f)=C_1(f)=2^n$.
Thus, improving the known $\Omega(2^n)$ lower bound requires going
above $\sqrt{C_0(f)C_1(f)}$.

To our knowledge, there is only one known lower bound for a total function which is better than $\sqrt{C_0(f)C_1(f)}$
(and no lower bounds for partial functions better than 
$\sqrt{\min(C_0(f), C_1(f)) N}$).
This is the $\Omega(N^{2/3})$ lower bound of Shi 
\cite{AS,Kutin,Ambainis03}
for element distinctness, a problem which has $C_0(f)=2$, $C_1(f)=N$ and 
$\sqrt{C_0(f)C_1(f)}=\Theta(\sqrt{N})$.
\comment{The second is lower bounds for binary search \cite{Ambainis99,HNS}.
There\footnote{In its usual formulation, binary search is
a problem with $N$-valued answer.
For example, we are given $x_1=\ldots=x_i=0$, $x_{i+1}=\ldots=x_N=1$ and would
like to find $i$. 
We can make it a problem with 0-1 valued answers by asking for $i\bmod 2$.
The known quantum lower bounds still apply.
The certificate consists of two bits: $x_i=0$ and $x_{i+1}=1$.}, 
$C_0(f)=C_1(f)=2$ but an $\Omega(\log n)$ lower bound is known.}
It uses methods quite specific to the particular problem
and cannot be easily applied to other problems.
\comment{
A more general approach is using Theorem 6 from \cite{Ambainis0}.
This is a generalization of Theorem \ref{AThm2} and, unlike Theorem \ref{AThm2},
it can give lower bounds better than $\Omega(\sqrt{C_0(f)C_1(f)})$.
An example of that is the lower bound for inverting a permutation \cite{Ambainis0}.
For this problem, $C_0(f)=C_1(f)=1$ but \cite{Ambainis0} shows a lower bound of
$\Omega(\sqrt{N})$.
Another similar example is the lower
bound for finding local minimum/maximum \cite{Aaronson1}.

Still, there are many more problems for which we suspect the query complexity to
be more than $\Omega(\sqrt{C_0(f)C_1(f)})$ but we cannot prove that.}
It would be very interesting to develop more methods of proving 
quantum lower bounds higher than $O(\sqrt{C_0(f)C_1(f)})$
for total functions or higher than $O(\sqrt{\min(C_0(f), C_1(f)) N})$
for partial functions.
\item 
{\bf Finding triangles.}
A very simple problem for which its true quantum complexity 
seems to exceed the $\Omega(\sqrt{C_0(f)C_1(f)})$ lower bound is as follows.
We have $n^2$ variables describing adjacency matrix of a graph.
We would like to know if the graph contains a triangle.
The best quantum algorithm needs $O(n^{1.3})$ queries \cite{Szegedy,MSS} 
an $\Omega(n)$ lower bound follows by a reduction from 
the lower bound on Grover's search \cite{Distinctness} or lower bound 
theorem of \cite{Ambainis0}.
We have $C_0(f)=O(n^2)$ but $C_1(f)=3$ (if there is a triangle, its three edges
form a 1-certificate), thus $\Omega(n)$ is the best lower bound
that follows from theorems in \cite{Ambainis0}.
We believe that the quantum complexity of this problem 
is more than $\Theta(n)$.
Proving that could produce new methods applicable to other problems
where quantum complexity is more than $O(\sqrt{C_0(f)C_1(f)})$
as well.

\end{enumerate}

{\bf Acknowledgments.}
Thanks to Scott Aaronson, Yaoyun Shi and Ronald de Wolf for
their comments about earlier versions of this paper.

\begin{appendix}

\section{Appendix: bounds using previous method}
\label{app:old}

In this section, we look at what bounds can be obtained for
$Q_2(f^d)$ for $f^d$ defined in section \ref{sec:4d}
using the previously known lower bound Theorem \ref{AThm2}.

\comment{ {\bf Attempt 1: $\Omega(2^d)$ lower bound.}
We start with a particular case of Theorem \ref{AThm2}
in which $R$ consists of $(x, y)$ with $x_i\neq y_i$ for
exactly one $i\in[N]$.

\begin{Theorem}
\label{AThm1}
\cite{Ambainis0}
Let $A\subset \{0, 1\}^N$, $B\subset\{0, 1\}^N$
be such that $f(A)=0$, $f(B)=1$ and 
\begin{itemize}
\item
for every $x=(x_1\ldots x_N)\in A$, there are at least $m$
values $i\in[N]$ such that
$(x_1\ldots x_{i-1}, 1-x_i, x_{i+1}, \ldots, x_N)\in B$,
\item
for every $x=(x_1\ldots x_N)\in B$, there are at least $m'$
values $i\in[N]$ such that
$(x_1\ldots x_{i-1}, 1-x_i, x_{i+1}, \ldots, x_N)\in A$.
\end{itemize}
Then, $Q_2(f)=\Omega(\sqrt{m m'})$.
\end{Theorem}

In the case of function $f^d$, 
this theorem gives a lower bound of $\Omega(2^d)$.
For that, we can just take $A=f^{-1}(0)$, $B=f^{-1}(1)$.
Since sensitivity of $f^d$ is $2^d$ on every input,
$m=m'=2^d$.
Also, the bound of theorem \ref{AThm1} cannot
be more the the maximum sensitivity of $f$ which is $2^d$.

{\bf Attempt 2: Using block sensitivity.}
In the more general Theorem \ref{AThm2},
we can flip blocks of variables instead
of single variables. Also, blocks do not
have to be disjoint (unlike in block sensitivity).
But, if they are non-disjoint, 
we have to account for non-disjointness
by having the maximum number of blocks
that share a variable in the denominator.
}
It can be verified that the block sensitivity of 
$f$ is $3$ on every input. 
By induction, we can show 
that this implies $bs_x(f^d)=3^d$ for every input $x\in\{0, 1\}^{4^d}$.
This makes it tempting to guess that we can achieve
$m=m'=3^d$ and $l=l'=1$ which would give a lower bound
of $\Omega(3^d)$.

This is not the case. If we would like to use Theorem \ref{AThm2}
with $l=l'=1$, we need two requirements simultaneously:
\begin{enumerate}
\item
For every $x\in A$, denote by $y_1, \ldots, y_{3^d}$ the elements of $B$ for which $(x, y_i)\in R$.
Then, the sets of variables where $(x, y_i)$ and $(x, y_j)$ differ must be
disjoint for all $i, j$, $i\neq j$.
\item
For every $y\in B$, denote by $x_1, \ldots, x_{3^d}$ the elements of $A$ for which $(x_i, y)\in R$.
Then, the sets of variables where $(x_i, y)$ and $(x_j, y)$ differ must be
disjoint for all $i, j$, $i\neq j$.
\end{enumerate}

If block sensitivity is $3^d$ on every input, 
we can guarantee the first requirement (by starting with $x\in A$ 
constructing disjoint $S_1$, $\ldots$, $S_{3^d}$ and putting
$(x, x^{S_i})$ into $R$). But, if the set $A$ only contains one $x$,
then $m'=1$ and the lower bound is $\Omega(\sqrt{3^d})$ which
is even worse than the previous one.

Therefore, we have to take larger set $A$.
This can break the second requirement. 
Let $x, z\in A$ and $y\in B$. Then, we could have 
$(x, y)\in R$ and $(z, y)\in R$. $x$ and $y$ would differ in a set
of variables $S_i$ which is one of $3^d$ disjoint blocks for $x$.
Similarly, $z$ and $y$ would differ in a set
$T_j$ which is one of $3^d$ disjoint blocks for $z$.
Now, there is no reason why $S_i$ and $T_j$ have to be disjoint! 
Block sensitivity guarantees that $S_i\cap S_j=\emptyset$ for every
{\em fixed} $x$ but it gives no guarantees about blocks for $x$ being
disjoint from blocks for $z$.

Similarly, if we start with $y\in B$, we can ensure the second
requirement but not the first.

The best that we could achieve with this approach was 
$m=m'=3^d$, $l=1$, $l'=2^d$, as follows.
Let $A=f^{-1}(0)$, $B=f^{-1}(1)$.
We inductively construct two sets of $3^d$ disjoint perfect 
matchings between inputs in $A$ and inputs in $B$.

The first set of matchings consists of ordered pairs $(x, y)$,
$x\in A$, $y\in B$.
For $d=1$, the first two matchings match each input $x\in A$ 
to the two inputs $y\in B$ that differ in exactly one variable.
The first matching is $(0011, 0001)$, $(0101, 1101)$, 
$(1100, 1110)$, $(1010, 0010)$, $(0100, 1100)$, $(1000, 0000)$, 
$(0111, 1111)$, $(1011, 1001)$.
The second matching matches each $x\in A$ to the other $y\in B$ 
which differs in exactly one variable. 
The third matching matches each $x\in A$ to $y\in B$ which
differs from $x$ in both variables that are sensitive for $x$.
This is the first set of 3 matchings.

The second set of matchings consists of ordered pairs  
$(y, x)$, $y\in B, x\in A$. The first two matchings are the same
as in the first set. The third matching matches each $x\in A$ to $y\in B$ which
differs from $x$ in both variables that are sensitive for $y$.

For $d>1$, we introduce notation $x^1$, $x^2$, $x^3$, $x^4$ and $\tilde{x}$ 
similarly to section \ref{sec:lem2}. The first $3^{d-1}$ matchings are constructed as 
follows. For each $x$, we find $\tilde{x}$. Then, we find $\tilde{y}$ such that
$(\tilde{x}, \tilde{y})$ belongs to the first matching in the first set. 
Let $i$ be the variable
for which $\tilde{x}_i\neq \tilde{y}_i$. In the $k^{\rm th}$ matching
($1\leq k\leq 3^{d-1}$), we match each $x\in A$ to $y\in B$ which is defined as follows:
\begin{itemize}
\item
If $j\neq i$, then $x^j=y^j$.
\item
$x^i$ is such that $(x^i, y^i)$ 
belongs to the $k^{\rm th}$ matching for $d-1$ levels (taking matchings from 
the first set if $f(x^i)=0$ and the second set if $f(y^i)=1$).
\end{itemize}
The second $3^{d-1}$ matchings are constructed similarly, except that we use
$\tilde{y}$ for which $(\tilde{x}, \tilde{y})$ belongs to the second matching of the first set.

To construct the last $3^{d-1}$ matchings, we take $\tilde{y}$ for which $(\tilde{x}, \tilde{y})$
belongs to the third matching. In $2\times 3^d+k^{\rm th}$ matching, we 
match $x$ with $y$ defined as follows.
\begin{itemize}
\item
if $\tilde{x}_i\neq \tilde{y}_i$, then $y^i$ is the input of length $x^i$ for which
$(x^i, y^i)$ belongs  to the $k^{\rm th}$ matching for $d-1$ levels. 
\item
if $\tilde{x}_i = \tilde{y}_i$, then $y^i=x^i$. 
\end{itemize} 

We then define $R$ as the set of $(x, y)$ which belong to one of the $3^d$ 
matchings we constructed. By induction, we show

\begin{Lemma}
For the first set of $3^d$ matchings, $m=m'=3^d$, $l=1$, $l'=2^d$.
For the second set of $3^d$ matchings, $m=m'=3^d$, $l'=1$, $l=2^d$.
\end{Lemma}

\proof
First, we prove $m=m'=3^d$. In the base case, we can just check 
that the matchings are distinct and, thus, every $x\in A$ or $y\in B$ is
matched to 3 distinct elements of the other set. In the inductive case,
consider an element $x\in A$ (or $y\in B$) and two elements $y_1\in B$ and $y_2\in B$ 
to which it is matched.
If $(x, y_1)$ and $(x, y_2)$ belong to two matchings in the same group of  $3^{d-1}$ matchings, 
then, by the inductive assumption $y_1^i\neq y_2^i$ and, hence, $y_1\neq y_2$.
If $(x, y_1)$ and $(x, y_2)$ belong to two matchings in different groups, then
$\tilde{y_1}\neq \tilde{y_2}$ implies $y_1\neq y_2$.

To prove $l=1$ (or $l'=1$ for the second set), we first observe that 
this is true in the base case. 
For the inductive case, we again have two cases.
If $(x, y_1)$ and $(x, y_2)$ belong to different sets of $3^{d-1}$ matchings, 
then, for each $i\in\{1, 2, 3, 4\}$, either $\tilde{x}_i=\tilde{y_1}_i$ or $\tilde{x}=\tilde{y_2}_i$.
This means that only one of $y_1$ and $y_2$ can differ from $x$ in a variable belonging to $x^i$.
If $(x, y_1)$ and $(x, y_2)$, we apply the inductive assumption to $(x^i, y^i_1)$ and $(x^i, y^i_2)$.

To prove $l'=2$ in the base case, we notice that, if $(x_1, y)$ and $(x_2, y)$ belong to the 
first and the second matching, then the pairs $(x_1, y)$ and $(x_2, y)$ cannot differ in the same variable.
In the inductive case, for every $i\in\{1, 2, 3, 4\}$, either $\tilde{x_1}_i = \tilde{y}_i$ or
$\tilde{x_2}_i=\tilde{y}_i$. If we have a variable $j$ such that $j\in\{(i-1)4^{d-1}+1, 
(i-1)4^{d-1}+2, \ldots, i\times 4^{d-1}\}$ and $\tilde{x_1}_i = \tilde{y}_i$, then 
$(x, y)\in R$ and $x_j\neq y_j$ means that $(x, y)$ belongs to either one of the second $3^{d-1}$
matchings or one of the last $3^{d-1}$ matchings. By applying the inductive assumption, there are at
most $2^{d-1}$ such $(x, y)$ in each of the two sets of $3^{d-1}$ matchings. 
This gives a total of at most $2\times 2^{d-1}=2^d$ such pairs $(x, y)$.
\qed

The weakness of Theorem \ref{AThm2} that we see here is that all
variables get treated essentially in the same way. For each $y\in B$,
different variables $y_i$ might have different number of $x\in A$ such that
$(x, y)\in R$, $x_i\neq y_i$. Theorem \ref{AThm2} just takes
the worst case of all of those (the maximum number).
Our weight schemes allow to allocate weights so that some
of load gets moved from variables $i$ which have lots of $x\in A$:
$(x, y)\in R$, $x_i\neq y_i$ to those which have smaller number of such $x\in A$.
This results in better bounds.

For the function of section \ref{sec:4d}, we get $\Omega(2.12..^d)$ by old
method and $\Omega(2.5^d)$ by the new method. For the two 
functions in section \ref{sec:other1}, the old method only gives
bounds that are lower than polynomial degree while the new method
shows that $Q_2(f)$ is higher than $\deg(f)$ for those functions as well.

\end{appendix}
\end{document}